\documentclass[oldversion]{aa}  
\usepackage{graphicx}
\usepackage{txfonts}
\usepackage{natbib}
\bibpunct{(}{)}{;}{a}{}{,} 
\usepackage{longtable,lscape}
\begin{document}
   \title{Gravitational waves from 3D MHD core collapse simulations}
   \author{S. Scheidegger
          \and
          T. Fischer
          \and
          S. C. Whitehouse
          \and
          M. Liebend\"orfer}
   \offprints{S. Scheidegger}
   \institute{Department of Physics, University of Basel, Klingelbergstrasse
82, 4056 Basel, Switzerland\\
             \email{Simon.Scheidegger@unibas.ch}}
   \date{Received August 30, 2007; Accepted July 2, 2008 }
%
 %
  \abstract
 {We present the gravitational wave analysis from rotating (model s15g) and nearly non-rotating (model s15h) 3D MHD core collapse supernova simulations at bounce and the first couple of ten  milliseconds afterwards. The simulations are launched from 15$M_{\odot}$ progenitor models stemming from stellar evolution calculations. Gravity is implemented by a spherically symmetric effective general relativistic potential. The input physics uses the Lattimer-Swesty equation of state for hot, dense matter and a neutrino parametrisation scheme that is accurate until the first few ms after bounce. The 3D simulations allow us to study features already known from 2D simulations as well as nonaxisymmetric effects. In agreement with  recent results we find only type I gravitational wave signals at core bounce. In the later stage of the simulations, one of our models (s15g) shows nonaxisymmetric gravitational wave emission caused by a low $T/|W|$ dynamical instability, while the other model radiates gravitational waves due to a convective instability in the protoneutron star. The total energy released in gravitational waves within the considered time intervals is $1.52\times10^{-7}M_{\odot}$ (s15g) and $4.72\times10^{-10}M_{\odot}$ (s15h).
Both core collapse simulations indicate that corresponding events in our Galaxy would be detectable either by the LIGO or Advanced LIGO detector.}
   \maketitle
%

\section{Introduction}

Gravitational wave astronomy could well make soon its first
observations as the running earth-based facilities such as LIGO\footnote{http://www.ligo.caltech.edu/}, VIRGO\footnote{http://www.ego-gw.it/},
GEO600\footnote{http://geo600.aei.mpg.de/}, TAMA\footnote{http://tamago.mtk.nao.ac.jp/} and AIGO\footnote{http://www.gravity.uwa.edu.au/} are up to reach the required high sensitivities. For a review on gravitational wave detection see e. g. \citep{lrr-2000-3}.
The successful observation of gravitational waves (GW) would be
a major breakthrough: it would open a new window to the universe,
allowing us to observe electromagnetically hidden regions \citep{1995pnac.conf..160T}.
One of the most promising gravitational wave sources is the stellar core collapse leading to
a supernova explosion. The observation of both, GW and the
neutrino signal, from a galactic supernova would reveal hitherto hidden
details about the explosion scenario and impose constraints on the
nuclear and weak interaction physics under conditions that can not
be obtained in terrestrial experiments.

As a core of a star of $ M \geq 8M_{\odot} $ reaches the end of its stellar evolution, it becomes gravitationally unstable as soon as thermonuclear burning produces a significant amount of iron group nuclei. Next to photo-disintegration, electron captures on free protons and nuclei reduce the mostly electron supported pressure, and the core starts to collapse eventually. The collapse continues until nuclear densities of $\sim2\times10^{14}$g/cm$^{3}$ are reached, depending on the equation of state (EoS). As soon as we enter this density regime, the core overshoots its equilibrium position and bounces back. A sound wave immediately forms and steepens into a shock wave that propagates outwards.
However, the shock stalls within
$\sim5$ ms after core bounce due to the large energy loss caused by
the dissociation of nuclei into free nucleons at a cost of $\sim8.8$ MeV
per nucleon and by neutrino emission connected to copious electron
captures on the emerging free protons. It continues to propagate
outwards to radii around \( 100-200 \) km as standing accretion shock.
Hereafter, i.e. some ms after the prompt explosion mechanism failed,
a delayed explosion mechanism by neutrino heating is thought to occur, e.g.
\citet{Bethe1990}.

The idea of reviving the stalled shock again via neutrino reactions behind and ahead of the shock has long been investigated as possible explosion mechanism \citep{Bethe1985,Janka2001a}. In the dissociated matter behind the shock, electron-flavour neutrino captures are the dominant heating reactions, while in the accreting matter ahead of the shock, electron-neutrino absorption and neutrino-nucleon scattering are the dominant interactions that preheat the infalling unshocked material \citep{Bruenn1991}. In addition, neutrino-electron scattering  and pair annihilation may contribute to the heating as well. However, the effects of pre-heating and neutrino-electron
scattering on the shock revival are insignificantly small compared to the effect from the capture
of neutrinos on free nucleons in the heating region below the shock. Equally as important as the neutrino heating is the neutrino cooling of matter that settles on the protoneutron star (PNS)
\citep{Janka2001}.

As a core collapse supernova is likely to show aspherical features
also close to its center \citep{2006Natur.440..505L}, where the matter assumes a high
density, a tiny fraction
of the released binding energy can be emitted via gravitational
radiation. The following features have been suggested as possible causes of
the asymmetries in the energy-matter distribution necessary to emit GW:
the rotational stellar core collapse, convection in the high-density
protoneutron star, nonaxisymmetric rotational instabilities, 
fluid instabilities in the lower-density hot mantle
surrounding it (possibly triggering neutron star oscillation modes),
and an anisotropic neutrino emission (for recent reviews
see e.g. \citep{2006RPPh...69..971K,Living Reviews in Relativity}). 

The understanding of GW emission from differentially rotating core collapse has evolved with time due to the improving input physics used in simulations.
In 2D axisymmetric computations by \citet{1982A&A...114...53M}  iron cores, a Newtonian hydrodynamics code and a tabulated finite temperature EoS were used. The results allowed to recognise the link between rotation and the efficiency of GW emission. 
After having taken into account more micro-physics in Newtonian gravity, such as electron capture on protons and a simplified neutrino transport scheme \citep{1991A&A...246..417M}, one was able to distinguish two different wave form  characteristics, later known as type I \& II. 
\citet{1997A&A...320..209Z} performed an extensive parameter study of a wide variety of models, using progenitors in rotational equilibrium, Newtonian gravity, a polytropic EoS, but neglecting electron capture and neutrino physics. The results lead to the introduction of type I, II \& III wave forms and their quantitative distinction in dependence of the stiffness of the EoS and rotation. 
A type I waveform is characterised
by a large amplitude peak at core bounce and subsequent damping
ring-down oscillations. It appears if the influence of the initial
angular momentum is small. In this case the core bounce occurs by the
stiffening of the equation of state and is pressure-dominated.
A type II signal occurs if the core bounce around
nuclear density is dominated by strong centrifugal forces; it has
several distinct peaks caused by multiple 'centrifugal' core bounces
with following coherent re-expansion phases of the inner core. The
type III signature is defined by a 'large' positive peak
at bounce followed by some smaller oscillations with very short
periods. It appears in the case of fast, pressure-dominated core collapse
due to very efficient electron capture.
Continued improvement of the input physics by using realistic equations of state or neutrino physics, by adding magnetic fields or incorporating GR effects, and by extending the dimensionality  from 2D to 3D, permitted to deepen and underline the qualitative and quantitative understanding of GW forms from rotational core collapse around bounce time \citep{Janka1996,1998A&A...332..969R,Fryer2002,Dimmelmeier2002,Kotake2003a,Kotake2004,M2004,2005PhRvD..71f4023D,Dimmelmeier07PhysRev,OttPhysRev07}.
In the most recent simulations by \citet{2007CQGra..24..139O} and \citet{Dimmelmeier07PhysRev} it was pointed out that realistic input physics (GR, a micro-physical EoS, approximative description of deleptonisation) might lead solely to type I wave forms.

The investigation of GW emission resulting from convectively driven small-scale aspherities inside the PNS, the 'hot-bubble' and anisotropic neutrino emission are still hampered by the requirement of computationally expensive neutrino transport in the postbounce evolution. Nevertheless, simulations that include accurate neutrino transport in axisymmetric models
were explored with respect to GW in \citet{M2004} based on state-of-the-art progenitor models, GR-corrections, and sophisticated equations of state. The postbounce phase is believed to provide for hundreds of milliseconds, or even longer than $1$ s, an interestingly large signal with a frequency distribution between $\approx300-1200$ Hz in case of convection, while the dominant frequencies lasting from anisotropic neutrino emission lie at about $\approx10$ Hz. Nonaxisymmetric dynamics has recently been investigated with neutrino physics approximations \citep{2007CQGra..24..139O}. In these computations it was possible to show without the addition of any seed perturbation that differentially rotating protoneutron stars are likely to become dynamically unstable at low $\beta=T/|W|$-values (ratio of rotational to gravitational energy), leading to strong narrow-band GW emission (\citet{ 2003ApJ...595..352S,2005ApJ...618L..37W,2005ApJ...625L.119O,2006AIPC..861..728S,2006ApJ...651.1068O}).

Another mechanism for gravitational wave emission in core collapse supernovae was proposed in  \citet{2006PhRvL..96t1102O}. It was pointed out that PNS core g-mode oscillations might be a major 
source of gravitational radiation. The release of GW in their simulations is highly correlated with the fundamental core g-mode, leading to a spectrum that peaks at twice the frequency of the aforementioned g-mode. 

In this article we present the gravitational wave analysis from two
3D magneto-hydrodynamics (MHD) core collapse simulations. The simulations are based on a nuclear equation
of state. The deleptonisation by electron capture and neutrino
emission are accurately parametrised up to about \( 5 \) ms after
bounce. Important general relativistic corrections to the gravitational potential
are taken into account as well. The simulations are launched from progenitor models
at the end of stellar evolution calculations
\citep{Woosley1995}.
One of the initial models (s15g) rotates moderately
fast at the onset of the collapse with $2\pi$ rad/s for the
central angular frequency, while the second model (s15h) rotates
more slowly as estimated in \citet{Heger2005}.
We perform contiguous simulations from collapse into the postbounce
phase and study three-dimensional effects on the emission of gravitational
waves. The GW-producing asymmetries in three-dimensional simulations
could be significantly different from the asymmetries present in axisymmetric
models. As only two other group have been performing three-dimensional
supernova models with a nuclear equation of state and neutrino physics
approximations for the prediction of gravitational wave signals \citep{Fryer2004,Dimmelmeier07PhysRev,OttPhysRev07}
our simulations provide an independent confirmation of the latter
results. Rather than augmenting the database with numerous GW pattern
templates, we aim to underline the importance of the neutrino physics
in supernova GW signal predictions and to encourage the search for
GW signals from Galactic core collapse supernova events.

The paper is organised as follows: In Sect. II we define the numerical
methods and input physics used in our supernova models. We also describe
the gravitational wave extraction formalism applied for the analysis
of our three-dimensional data sets. Section III is devoted to the
resulting gravitational waves, their properties and the possible detectability
of signals. In Sect. IV we discuss the uncertainties in the numerical
models and try to separate the robust features from details that are
subject to change in future improved models. A conclusion is given
in Sect. V.

\section{Method}

\subsection{Description of the hydrodynamical models s15g and s15h}

Only few recent multi-dimensional collapse simulations made the effort
to include neutrino physics. These schemes are either very computationally
expensive \citep{Buras2003,Dessart2006} or rely on simplifications
of the neutrino transport and its micro-physics \citep{Kotake2003a,2004ApJ...601..391F}.
The complete Boltzmann neutrino transport equation can only be solved
in spherical symmetry (\citep{Mezzacappa2005} and references therein).
Hence, three-dimensional hydrodynamical models must rely on neutrino
transport approximations.

A simple and computationally efficient parametrisation of the deleptonisation
in the collapse phase can be derived from a tabulation or fit of the
electron fraction \( Y_{e} \) as a function of density. Spherically
symmetric models with Boltzmann neutrino transport \citep{Liebend2005}
show that the \( Y_{e} \) in different layers in the homologously
collapsing core follow a quite similar deleptonisation trajectory,
i.e. reach similar \( Y_{e} \) values as a function of density. As
changes in \( Y_{e} \) can only be due to electron captures, it is
possible to deduce corresponding changes in entropy and to calculate
the neutrino stress from the emitted neutrinos \citep{Liebendorfer2005}.
\begin{figure}
   \centering
    \includegraphics[width=8.8cm]{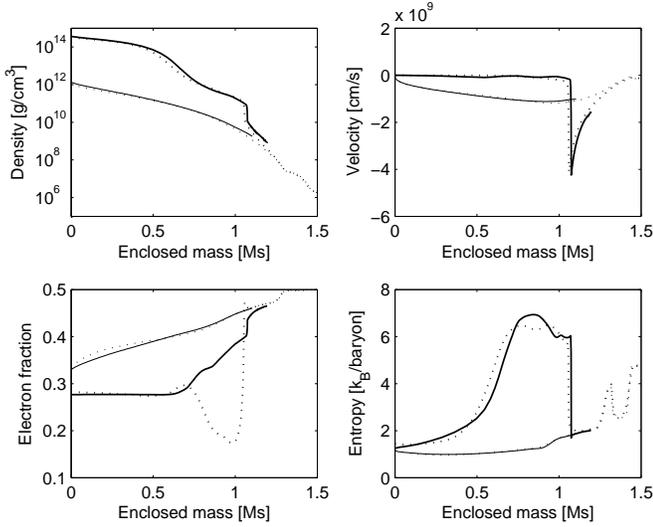}
      \caption{Comparison of the almost non-rotating 3D model s15h 
    with the spherically symmetric model G15, which is based on general
    relativistic three-flavour Boltzmann neutrino transport \protect\citep{Liebend2005}.
    From the upper left to the lower right we compare as a function of enclosed
    mass: the density-, the velocity-, the \( Y_e \)-, and the entropy profiles. The
    solid lines show the results of model s15h and the dotted lines the results of
    model G15. The thin lines represent a time instance at \( 5 \) ms before bounce
    and the thick lines represent a time instance at \( 5 \) ms after bounce.
    Excellent agreement is found in all four quantities---with one
    exception: The parametrised neutrino leakage scheme cannot model the
    neutrino burst. The neutrino burst causes a prominent
    \( Y_e \)-dip and additional cooling in the G15 data.}
         \label{fig1.eps}
   \end{figure}
Figure \ref{fig1.eps} shows a comparison of the 3D parametrised run
s15h, launched from an almost non-rotational progenitor model, with
a spherically symmetric general relativistic model G15 that is based
on general relativistic Boltzmann neutrino transport \citep{Liebend2005}.
The 1D results of core collapse are accurately reproduced by the 3D
run. The parametrised neutrino physics presents a significant improvement
with respect to adiabatic simulations and may even rival with neutrino
transport schemes that neglect neutrino-electron scattering. However,
the accuracy breaks down with the launch of the neutrino burst at
a few milliseconds after bounce. With the currently implemented scheme,
accretion flows in the postbounce phase deleptonise only down to \( Y_{e}\sim 0.3 \)
instead of \( Y_{e}\sim 0.15 \). Neutrino heating is neglected altogether.
This turns the quantitative model of the collapse phase into a qualitative
model of the postbounce phase.

The hydrodynamics of the 3D simulations is based on a simple and fast
cosmological MHD code \citep{Pen2003} which has been parallelised,
improved and adapted to the supernova context. A realistic equation
of state \citep{Lattimer1991} is used and the monopole term of the
gravitational potential is implemented by a spherically symmetric
mass integration which includes general relativistic corrections \citep{Marek2006}.
The 3D computational domain consists of a central cube of \( 600^{3} \)
km\( ^{3} \) volume, treated in equidistant Cartesian coordinates
with a resolution of \( 1 \) km. The 3D code has buffer zones that
require at each time step a prescription of the conditions in the
immediate neighborhood of the 3D computational domain. In order to
obtain these conditions, we embed the 3D computational domain in a
larger spherically symmetrical computational domain that is treated
by a spherically symmetric hydrodynamics code 'Agile' \citep{Liebend2002}.
After each time step, the buffer zones at the boundary of the 3D hydrodynamics
code are filled with the current conditions of the spherically symmetric
solution. As long as the infall velocity at the boundary is subsonic,
we spherically average the conditions in the 3D code and feed them
back to the mass shells of Agile that are enclosed by the 3D domain.
Once the infall velocities become supersonic, this step can be omitted
because no hydrodynamic signal can leave the 3D domain anymore before
the shock passes the boundary after the onset of an explosion.

For both of our models we use a \( 15 \) M\( _{\odot } \) progenitor
star of \citep{Woosley1995} as initial model. As the rotation rates
of inner stellar cores are not very well-known, we assign angular
momentum according to a simple parametrisation with a shellular quadratic
cutoff at \( 500 \) km radius. The angular momentum is assumed to
be conserved until the infalling layers enter the 3D computational
domain. For one model, which we name s15h, we set an initial angular
velocity of \( \Omega =0.3 \) rad/s as obtained in a stellar evolution
model that takes account of the effects of rotation and magnetic fields
\citep{Heger2005}. Because this value represents a rather slow rotation,
we performed an alternative model, s15g, with an initial angular velocity
of \( \Omega =2\pi  \) rad/s. This is still a slow rotation rate
compared to values assumed in many parametrised studies of the GW
signal, but it allows for a clear distinction from model s15h. The
initial \( \beta _{i}=T/|W| \) comports \( 0.26\% \) (s15g) and
\( 0.00059\% \) (s15h), respectively. Both models collapse in a similar
manner to spherically symmetric simulations. After bounce, the declining
strength of the bounce-shock results in a negative entropy gradient
and the protoneutron star becomes convectively unstable on a short
time scale. The simulation s15g was carried out until about
$70$ ms postbounce, whereas s15h run $\sim100$ ms postbounce.

We have set the magnetic fields to the values suggested in \citet{Heger2005}.
The advection of the magnetic field into the boundary
of the 3D computational domain is technically very delicate because
the field lines assigned to the buffer zones have to ensure a vanishing
divergence at the transition to the slightly evolved magnetic field
on the interior of the boundary. We found a quite pragmatic solution
where the field is first assigned according to above analytical setup
of a divergence-free initial field. In a second step, the divergence
at the boundary is analysed and corrected such that the transition
is divergence-free. In any case, due to its weakness, the magnetic
field does not influence the hydrodynamical evolution noticeably during
the early postbounce evolution.


\subsection{Extracting gravitational radiation in three dimensions}

Throughout this article we work in cgs-units and use the following values for the speed of light, c$=2.997\cdot10^{10}$ cm s$^{-1}$, the gravitational constant, G$=6.672\cdot10^{-8}$cm$^{3}$g$^{-1}$s$^{-2}$, and the parsec, $1$ pc $=3.086\cdot10^{18}$ cm. 
We do not assume any symmetry. In this case, the gravitational wave field $h_{ij}^{TT}$ can be decomposed into two orthogonal polarisations with amplitudes $A_{+}$ and $A{\times}$, see e.g.  \citep{1973grav.book.....M,1990ApJ...351..588F}:
\begin{equation}
 h_{ij}^{TT}(\textbf{X},t)=\frac{1}{R}(A_{+}e_{+}+A_{\times}e_{\times}),
\end{equation}
where $R$ is the distance to the source.
In spherical coordinates, the unit polarisation tensors are given by:
\begin{equation}
 e_{+}=e_{\theta}\otimes e_{\theta}-e_{\phi}\otimes e_{\phi}
\end{equation}
and 
\begin{equation}
 e_{\times}=e_{\theta}\otimes e_{\phi}+e_{\phi}\otimes e_{\theta}.
\end{equation}

In the slow-motion limit \citep{1973grav.book.....M,1990ApJ...351..588F} the amplitudes $A_{+}$ and $A_{\times}$ are given by linear combinations of the second time derivative of the transverse traceless mass quadrupole tensor:
\begin{eqnarray}
 A_{+} & = & \ddot{t}_{\theta\theta}-\ddot{t}_{\phi\phi} \\
 A_{\times} & = & 2\ddot{t}_ {\theta\phi}.
\end{eqnarray}
In the Cartesian orthonormal basis, the quadrupole tensor is given by  
\begin{equation}
t_{ij}^{TT}=\frac{G}{c^4}\int dV \rho\left(x_{i}x_{j}-\frac{1}{3}\delta_{ij}r^2\right).
\label{equ:1}
\end{equation} 
Note that its time derivatives are evaluated at a retarded time. However, there exist shortcomings related to numerical high-frequency noise and the $r^2$ momentum-arm which make the performance of a direct evaluation of Eq. (\ref{equ:1}) poor, as discussed in \citep{1990ApJ...351..588F}. Therefore, we use an alternative post-Newtonian expression from \citet{1990MNRAS.242..289B}, where the  second order time derivatives of the quadrupole moment are transformed into hydrodynamical variables which are known from the core collapse simulation.  
\begin{equation}
\ddot{t}_{ij}^{TT}=\frac{G}{c^4}\int dV \rho\left(2v_{i}v_{j}-x_{i}\partial_{j}\Phi-x_{j}\partial_{i}\Phi\right),
\label{equ:2}
\end{equation} 
where $\Phi$ is the gravitational potential. 
This expression allows the evaluation of the wave field from data at only one time instance; moreover, the integral has a compact support. Note, however, that this formula is not gauge invariant and only valid in the Newtonian slow-motion limit. Nevertheless, it was shown that this approximation is sufficiently accurate when compared to other approaches \citep{2003PhRvD..68j4020S}.

The polarisation modes can explicitly be obtained from a coordinate transformation, for example for the $\theta\theta$-component:
\begin{equation}
t_{\theta\theta}=t_{ij}^{TT}\frac{\partial x^{i}}{\partial\theta}\frac{\partial x^{j}}{\partial\theta}.
\end{equation}
This leads to the following non-vanishing components \citep{Oohara1997}:
\begin{eqnarray*}
t_{\theta\theta}& = & (t_{xx}^{TT}\cos^{2}\phi+t_{yy}^{TT}\sin^{2}\phi+ 2t_{xy}^{TT}\sin\phi\cos\phi)\cos^{2}\theta  \\
		& + & t_{zz}^{TT}\sin^{2}\theta-2(t_{xz}^{TT}\cos\phi +t_{yz}^{TT}\sin\phi)\sin\theta\cos\theta \\
t_{\phi\phi} &=& t_{xx}^{TT}\sin^{2}\phi+t_{yy}^{TT}\cos^{2}\phi-2t_{xy}^{TT}\sin\phi\cos\phi\\
t_{\theta\phi}&=&(t_{yy}^{TT}-t_{xx}^{TT})\cos\theta\sin\phi\cos\phi+ t_{xy}^{TT}\cos\theta(\cos^{2}\phi\\
              &-&\sin^{2}\phi)+t_{xz}^{TT}\sin\theta\sin\phi-t_{yz}^{TT}\sin\theta\cos\phi.
\end{eqnarray*} 
We evaluate below for simplicity only the gravitational wave amplitudes for $\theta=\phi=0$ (denoted below with the subscript $I$)
\begin{eqnarray}
A_{+I} & = & \ddot{t_{xx}}-\ddot{t_{yy}} \\
A_{\times I} & = & 2\ddot{t_{xy}}
\end{eqnarray}
and for $\theta=\frac{\pi}{2}$, $\phi=0$ (denoted as II)
\begin{eqnarray}
A_{+II} & = & \ddot{t_{zz}}-\ddot{t_{yy}} \\
A_{\times II} & = & -2\ddot{t_{yz}}.
\end{eqnarray}
We notice that $A_{+II}$ corresponds to the non-vanishing quadrupole amplitude $A_{20}^{E2}$ of axisymmetric models in the multipole expansion of the radiation field \citep{1980RvMP...52..299T}.

The energy carried away by gravitational radiation can be calculated by the following expression: 
\begin{eqnarray}
E_{GW} & = & \frac{c^3}{5G}\int [\frac{d}{dt}(I_{ij}-\frac{1}{3}\delta_{ij}I_{ll})]^2 dt
\label{equ:3} \\
&=&\frac{2c^3}{15G}\int dt[\dot{I_{xx}}^2+\dot{I_{yy}}^2 + \dot{I_{zz}}^2\nonumber\\
&-& \dot{I}_{xx}\dot{I}_{yy}- \dot{I}_{xx}\dot{I}_{zz}-\dot{I}_{yy}\dot{I}_{zz}\nonumber\\
&+&3(\dot{I}_{xy}^2+\dot{I}_{xz}^2+\dot{I}_{yz}^2)], \nonumber
\end{eqnarray}
where $I_{ij}=\ddot{t}_{ij}$. 

Whenever the question is raised whether the emitted signal of a source would possibly be detectable by earth-based facilities, such as e. g. LIGO, one is in particular interested in the detector dependent characteristical frequencies $f_{c}$ and amplitudes $h_{c}$ (calculated according to Eq. (31) in \citep{1989thyg.book.....H}) and the 'signal-to-noise ratios' (SNR). 
We estimate the signal-to-noise ratios for optimal filtering searches as follows
\citep{1998PhRvD..57.4535F,M2004}:
\begin{eqnarray}
 \left(\frac{S}{N}\right)^{2}_{optimal filter}&=&4\int \frac{|\hat{h}(\nu)|^{2}}{S_{h}(\nu)}d\nu \\
\label{equ:4}
&=&4\int \left[\frac{|\hat{h}(\nu)|\nu^{1/2}}{h_{rms}(\nu)}\right]^2d\log\nu,
\label{equ:5}
\end{eqnarray}
where $\hat{h}(\nu)$ is the Fourier transform of the gravitational wave amplitude,
\( h(\nu) = h_{+}(\nu) + h_{\times}(\nu) \), and $S_{h}(\nu) [1/Hz]$ is the power spectral density of strain noise in the detector and $h_{rms} [Hz^{-1/2}]\equiv\sqrt{S_{h}(\nu)}$ stands for the rms noise level of the detector at a given frequency $\nu$.
The Fourier spectra were normalised by using Parseval's Theorem.

Below, we determined the signal-to-noise ratios for an optimal orientation of detector and source. As detector sensitivity we used the present performance of one single LIGO instrument and the improved, possible future performance of an Advanced LIGO (AdvLIGO) detector \citep{Shoemaker:2007}. Note that Advanced LIGO possesses several adjustable frequency responses. While the 'burst' selection provides a broad range of nearly maximum sensitivity (optimal for model s15g), a 'nsns'-tuned instrument is likely to be used for sources that radiate at lower frequencies (model s15h).


\section{Results}
\begin{figure}
   \centering
    \includegraphics[width=8.8cm]{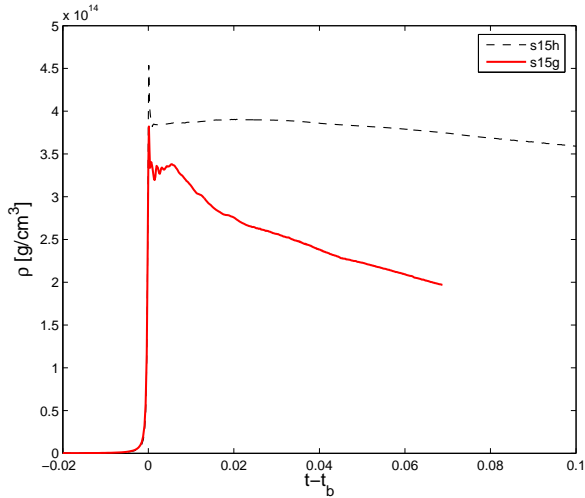}
	\caption{Time evolution of the central densities of the models s15g and s15h. At core bounce the following values are assumed:
	$\rho_{c}=3.8 \times
	10^{14}$ g/cm$^3$ for model s15g and $\rho_{c}=4.5 \times 10^{14}$ g/cm$^3$  for model
	s15h. The density decrease in the postbounce phase is most likely due to numerical dissipation of steep density gradients in the neutron star.}
	     \label{fig2.eps}
   \end{figure}
\begin{figure}
   \centering
    \includegraphics[width=8.8cm]{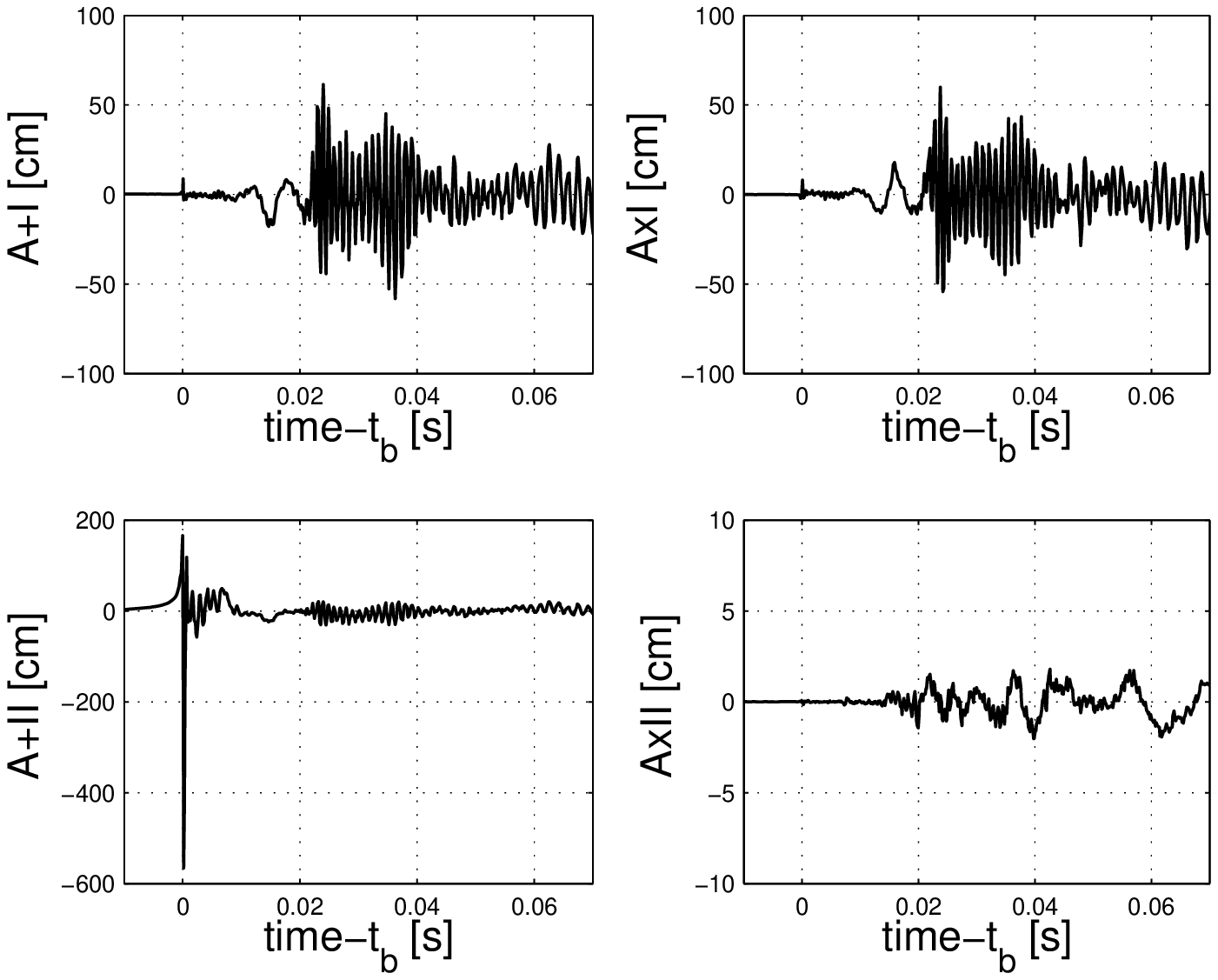}
	\caption{Time evolution of the quadrupole amplitudes A+I, AxI, A+II and AxII for model s15g.}
	     \label{fig3.eps}
   \end{figure}
\begin{figure}
   \centering
    \includegraphics[width=8.8cm]{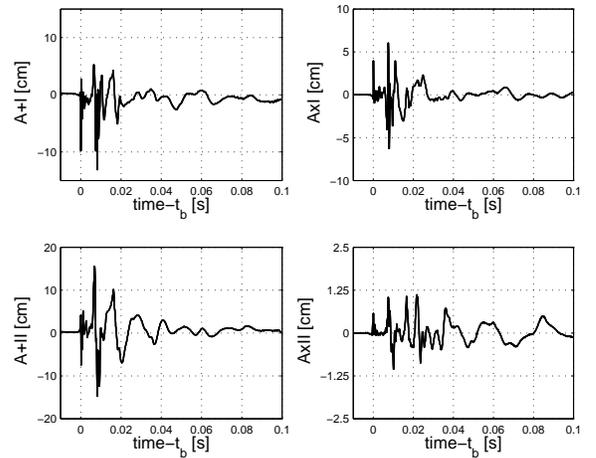}
	\caption{Time evolution of the quadrupole amplitudes A+I, AxI, A+II and AxII for model s15h.}
	     \label{fig4.eps}
   \end{figure}
The computed quadrupole wave amplitudes of models s15g and s15h are displayed in Figs. \ref{fig3.eps}, \ref{fig4.eps} and \ref{fig6.eps}. A quantitative overview of key properties and results is given in Table \ref{table:1} and Fig. \ref{fig5.eps}. Detector dependent quantities are summarised in Table \ref{table:2} and Fig. \ref{fig8.eps}.
\begin{table*}
\caption{Summary of GW related quantities for the models s15g and s15h. \protect\thanks{
$\beta=T/|W|$ is a convenient measure of the average rotation rate, which is calculated by taking the ratio of rotational to gravitational energy. $E_{gw}$ is the total energy released in gravitational radiation and $f_{b}$ is the frequency of strongest emission at bounce. Finally, we present the maximum amplitudes at different stages of their time-evolution in polar (I) and equatorial (II) direction. The subscript $i$ stands for $initial$, while $b$ and $pb$ stand for $bounce$ and $post-bounce$.}}
\label{table:1}
\centering
\begin{tabular}{cccccrrrr}
Model & $\beta_{i} $ & $\beta_{b}$ & $E_{GW}[M_{\odot} c^2]$ & direction & $|A_{+,b}|[cm]$ & $|A_{\times,b}|[cm]$ & $|A_{+,pb}|[cm]$ & $|A_{\times,pb}|[cm]$\\
\hline \hline\\
 s15g & $0.26\times10^{-2}$ & $5.2\times10^{-2}$ & $1.52\times10^{-7} $ & I & $9$ & $8$ & $62$ &
$59$ \\
        &         &                      &              & II&$566$&$<1$ & $30$ &
$2$  \\
\hline \\
 s15h & $0.59\times10^{-5}$  & $1.7 \times10^{-4}$   & $4.72\times10^{-10}$ & I & $10$ & $4$ &  $13$ &
$6$  \\
        &         &                      &              & II& $8$& $<1$& $15$ &
$1$ \\
\hline
\end{tabular}
\end{table*}

\subsection{Model s15g}

Model s15g undergoes a rotational core collapse as we assume an initial central angular velocity of $\Omega=2\pi$ rad/s. During the stage of contraction, the core spins up massively while becoming oblate. The collapse gets abruptly halted due to the stiffening of the EoS above nuclear densities. Afterwards the core rebounds and drives a hydrodynamical shock wave outwards. These conditions give rise to strong time-dependent variations in the physical quantities that affect the quadrupole tensor (see equ.\ref{equ:2}).
Its behaviour is then directly reflected by the model's gravitational wave signature. 
Since the core collapse proceeds nearly axisymmetrically, the only GW amplitude considerably driven by the rotationally induced large-scale asymmetries is A$_{+II}$. It exceeds the other s15g-wave trains A$_{+I}$,A$_{\times I}$ and A$_{\times II}$ by 1-2 orders of magnitude, as one can see in Fig.  \ref{fig3.eps} for times around bounce. The initial GW signal is emitted just before core bounce, when the rapid infall of matter and the spin-up of the core is dominant. It shows a prebounce rise.  
Then, the instantaneous slowdown of matter at core bounce leads to a prominent negative peak, which is followed by a ring-down behaviour that lasts for the first few ms postbounce. This generic type I wave characteristics is displayed in the lower left panel of Fig. \ref{fig3.eps}.
Performing a Fourier transform of the GW signal around bounce ($-5$ ms $<t<5$ ms), we find a spectrum with a very narrow bandwidth peaking around $893$ Hz. This is in good agreement with the recent findings from \citet{M2004}, but more than $150$ Hz higher than in  \citep{Dimmelmeier07PhysRev,2007CQGra..24..139O}.
It has been suggested that the difference stems from the fact that full relativistic calculations shift the bounce spectrum to lower frequencies in comparison to the ones using an effective gravitational potential (\citet{Dimmelmeier:2007}).

After this first and predominantly axisymmetric stage of GW-emission, the occurrence of a low $T/|W|$-instability revives the gravitational wave signal again around $\approx 20$ms post-bounce (\citet{ 2003ApJ...595..352S,2005ApJ...618L..37W,2005ApJ...625L.119O,2006AIPC..861..728S,2006ApJ...651.1068O,2007CQGra..24..139O}).
Low $T/|W|$ dynamical instabilities are triggered in differentially rotating systems such as neutron stars in situations where the patten speed $\sigma_{p}=\sigma/m$ of an unstable mode $m$ matches the local angular velocity at a point in the star (see Fig. \ref{fig7.eps}), commonly called \textit{corotation point} (the modes are, as in \citet{2005ApJ...618L..37W}, assumed to behave harmonically as $\exp[-i(\sigma t -m\phi)]$, where $\sigma$ is the mode's eigenfrequency). It permits the azimuthal fluid modes to amplify.
This non-axisymmetric process yields a quasi-periodic GW signal with a rather constant time-variation, leading to a narrow-band emission at $905$ Hz which lasts until the end of our simulation, as one can see particularly in the upper panels of Fig. \ref{fig3.eps} for times t $>20$ ms.
The analysis method we use to observe the growth of nonaxisymmetric structures
decomposes the density at a fixed radius \textit{R} and constant \textit{z}-component into its azimuthal Fourier components as done before e.g. in ref. \citep{2006ApJ...651.1068O,2007CQGra..24..139O}):

\begin{equation}
\rho (R,z,\phi) =\sum_{m = -\infty}^{\infty}C_{m}(R,z)e^{im\phi}, 
\end{equation}

where the complex Fourier amplitudes are defined by 

\begin{equation}
C_{m} = \frac{1}{2\pi} \int_{0}^{2\pi}\rho(R,z,\phi)e^{-im\phi}d\phi
\end{equation}

\begin{figure}
   \centering
    \includegraphics[width=8.8cm]{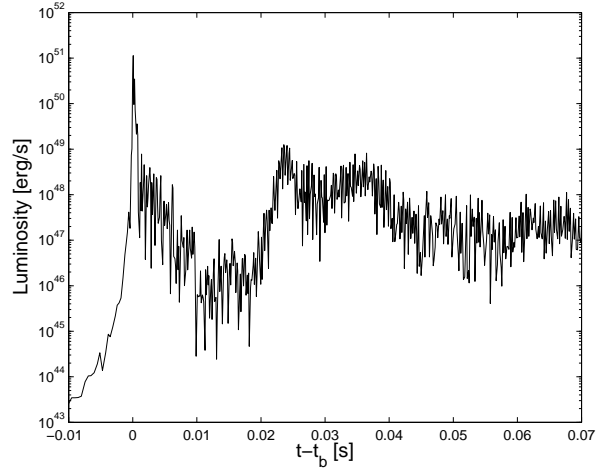}
	\caption{Time evolution of the GW luminosity in model s15g. In this  70 ms
	window, about $90\%$ of the energy emitted in GW stems from the A+II contribution
	around bounce.}
	     \label{fig5.eps}
   \end{figure}

\begin{figure*}
   \centering
    \includegraphics[width=18cm]{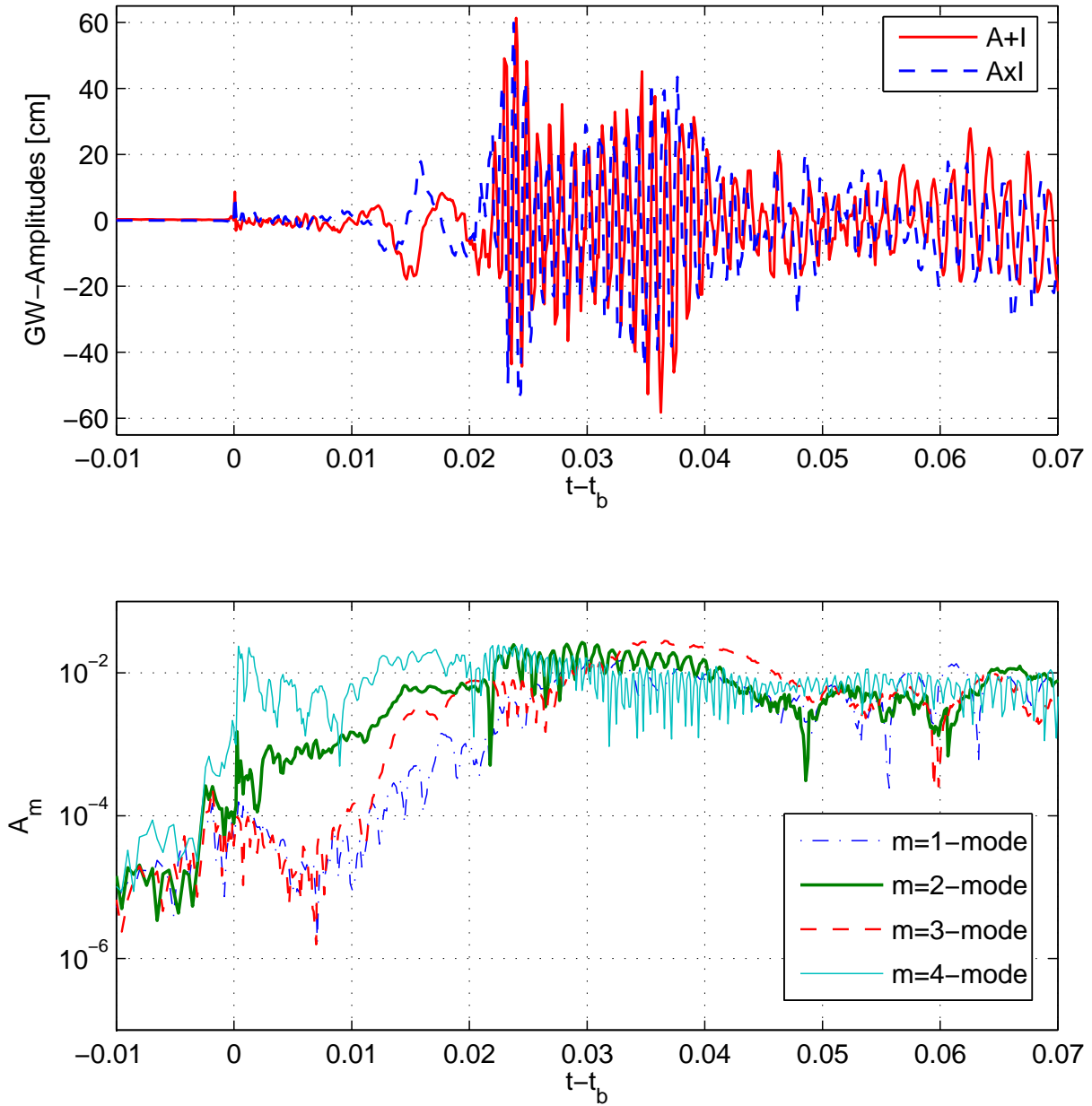}
	\caption{The upper panel displays the GW emission of the \textit{$A_{+}$} and the \textit{$A_{\times}$}-amplitude along the pole. The lower panel shows the normalized mode amplitudes $A_{m}$ for $m=\{1,2,3,4\}$ extracted at a radius of $20$ km. Note the sudden onset of the nonaxisymmetric GW-signal along the pole as soon as the $m=2$ mode amplitude reaches approximately the absolute size of the $m=4$ grid mode background. Notice further that $A_{+}$ and $A_{\times}$ oscillate at the same frequency of $905Hz$, phase shifted by $\pi/2$, as one would expect from a rotating bar.}
	     \label{fig6.eps}
   \end{figure*}

\begin{figure}
   \centering
    \includegraphics[width=8.8cm]{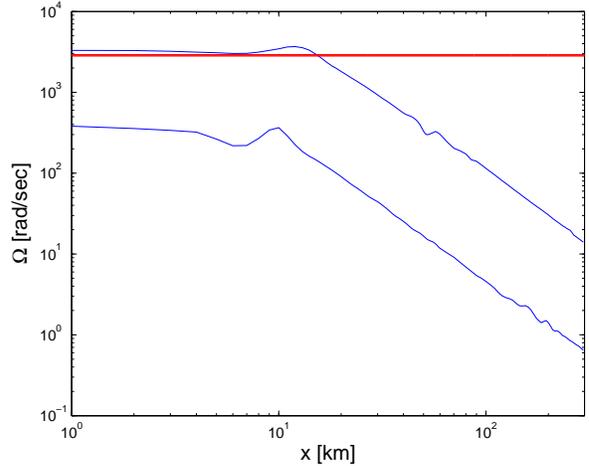}
	\caption{Angular velocity profiles of model s15g (upper profile) and s15h (lower profile) along the positive x-axis in the equatorial plane at $t-t_{b}\approx6$ms (S15g) and $t=t_{b}$ (s15h), respectively. The $m=2$ mode pattern speed of s15g is given by the horizontal line. Note first that both models' innermost $\approx 10$ km are nearly in perfect solid body rotation and secondly that model s15g is in corotation at a radius of about $10$ km, spinning with some $2850$ [rad/sec].}
	     \label{fig7.eps}
   \end{figure}

In Fig. \ref{fig6.eps} the normalized mode amplitudes $A_{m}=|C_{m}|/C_{0}$ are monitored to measure the growth of unstable modes.
In our model s15g we find $m=\{1,2,3\}$-modes being triggered, with the so-called $m=2$ bar-mode growing fastest. Further, we state that the $m=\{1,2,3\}$ modes all possess the same pattern speed. The close relation between the $m=2$ bar-mode instability and the emission of gravitational waves can be seen in the following features: First, the sudden onset of GW emission along the pole, which must be completely due to nonaxisymmetric dynamics, coincides with the amplitude of the $m=2$ mode reaching approximately the absolute amplitude of the $m=4$ mode caused by the grid. Secondly, the dominant frequency of emission corresponds perfectly to the eigenfrequency of the $m=2$-mode. Finally, the two GW-polarisations $+$ and $\times$ are phase shifted by $\pi/2$, as one would expect of a perfect, monochromatic GW-source such as a rotating bar.
These findings in the context of the low $T/|W|$ instability and supernova dynamics stand in remarkable agreement with the recent ones of \citet{2007CQGra..24..139O}. For low $\beta$-unstable models similar to s15g, they found narrow-band GW emission at $\approx 920-930$ Hz. The main difference to our calculations is the point that the dominant mode which was found in those computations was the $m=1$-mode. 
As a closing remark to model s15g we state that the time evolution of the energy emitted by gravitational radiation (see Fig.\ref{fig5.eps}) fits the behaviour of the waves. It demonstrates a large peak around bounce at $1.3\times10^{51}$ erg/s followed by a ringdown and an oscillating renaissance at about $10^{47}-10^{48}$ erg/s for times $t>20$ ms.

\subsection{Model s15h}

The slow-rotating model s15h undergoes a quasi-spherically symmetric core collapse, consequently showing fairly weak type I amplitudes around bounce (see Fig. \ref{fig4.eps}).
On the other hand, the marginally present centrifugal forces allow a core bounce at higher central densities ($4.5\times10^{14}$ g/cm$^{3}$) than in the previously discussed model ($3.8\times10^{14}$  g/cm$^{3}$) as one can see in Fig. \ref{fig2.eps}. 
The general dependence of A+II on rotationally induced asymmetries gets obvious in the following feature: while the A+II amplitude from s15h at bounce is more than one order of magnitude smaller as in s15g,
all other amplitudes, namely A+I, AxI and AxII are of the same order of magnitude as their counterparts from model s15g, emphasising a general feeble subordination to rotationally caused aspherities around bounce.

Prompt convection, based on the negative entropy gradient left behind the stalling shock starts just a bit before $10$ ms after bounce.
It leads to a convectively driven rise of the GW signal; its characteristic is dominated by the stochastic process of convective mass motion: neither a clear signal type nor any correlation between the two polarisations as can be found here, as one expects from this kind of matter motion. 
However, there is a peak in the wave train at $\approx7$ ms, best visible in A+II, which we could
not attribute to a physical feature of core collapse. It is most likely the result of a grid alignment
effect at a radius of 30-50 km. When the fluid first breaks its spherical symmetry in this region, this
leads to a numerical quadrupole moment in the fluid aligned with the main coordinate axes.
The overall spectra arising from the s15h-GW amplitudes is qualitatively and quantitatively rather different to the one of s15g: it ranges from some Hz to about $1000$ Hz with major contributions below $500$ Hz, clearly dominated by post-bounce convective motions (see Fig. \ref{fig8.eps}).
The signal lasting from later times ($t>20$ ms) mainly takes place at a frequency range from roughly $250$ Hz down to about about $10$ Hz. This broad band emission of GW mirrors the wide spread and inhomogeneous velocity-distribution of the convective motion. 
The total energy emitted in GW is nearly three orders of magnitude lower than in the previously discussed model (see Table \ref{table:1}), and the max luminosity reaches only $\approx{3\times10^{47}}$ erg/s at core bounce.
The net size of our convectively driven amplitudes (some centimetres) as well as the frequency band of emission and the total amount of energy released in gravitational radiation fit qualitatively well the results of \citet{M2004}, where they performed 2D core collapse simulations of slowly rotating progenitor models with similar input physics.

  
\subsection{Detectability}

\begin{figure*}
   \centering
    \includegraphics[width=18cm]{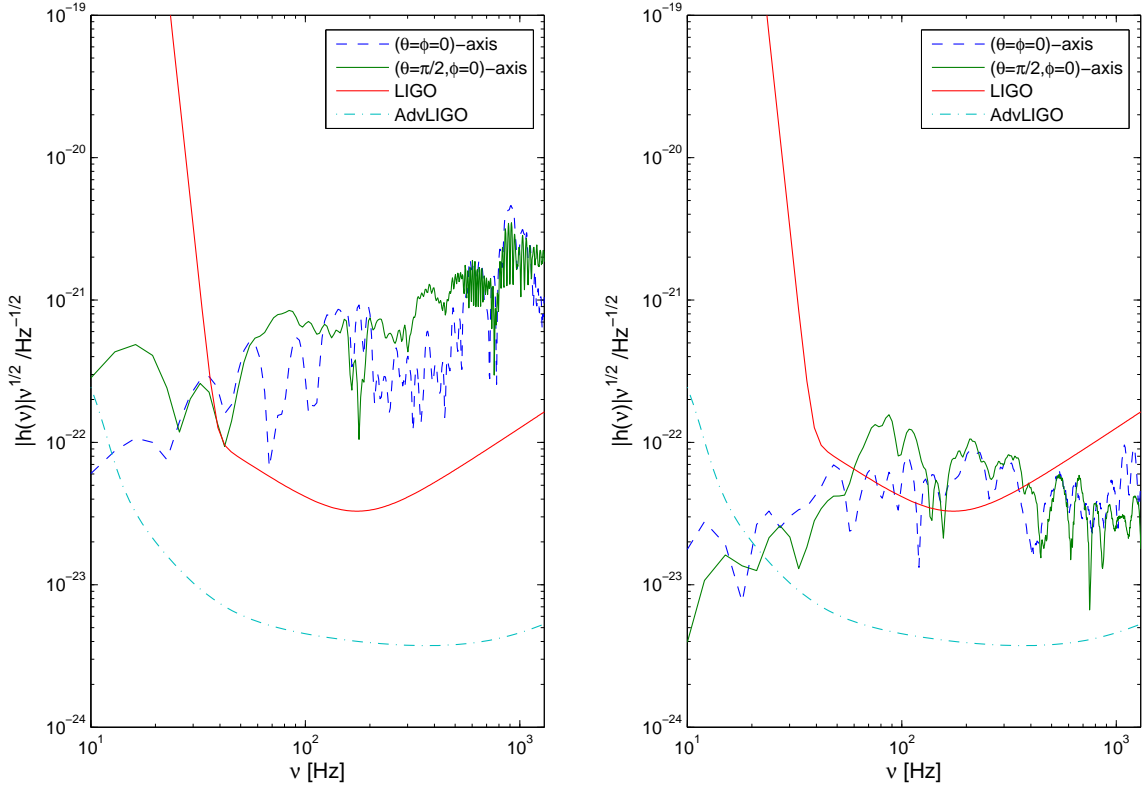}
	\caption{Spectral energy distribution of the GW signals at a distance of $10$ kpc in comparison with the present LIGO strain sensitivity and the possible performance
of Advanced LIGO (broadband tuning). The strain sensitivity curves have kindly been provided by \protect\citet{Shoemaker:2007}. The left panel shows the result for the rotational model s15g and the right panel shows the result for the non-rotational model s15h. Optimal orientation between source and detector is assumed and the polarizations are combined according to Eq. (\protect\ref{equ:4}). The dashed lines show the GW signal for an observer on the rotational axis of the source, the solid lines show the GW signal for an observer in the equatorial plane. Note model s15g's signal enhancement of the in the equatorial plane around $900$ Hz finds its physical origin in the core bounce, while the polar signal around the same frequency is dominated by the $m=2$ dynamical instability. Note also the spectral energy distribution of s15h peaks in the region of maximal detector sensitivity. This would probably allow to observe the post-bounce GW-emission from initially nearly spherical symmetric core collapse supernovae.}
	     \label{fig8.eps}
   \end{figure*}
As in \citet{M2004}, in Fig. \ref{fig8.eps} we compare the quantity $|\hat{h}(\nu)|\nu^{1/2}$ in Eq. (\ref{equ:5}), evaluated at a Galactic distance of $10$ kpc for a single LIGO detector's $h_{rms}$.
\begin{table}
\caption{Summary of detector-dependent quantities. \protect\thanks{
We assume the sources to be located at a distance of $10$ kpc as measured by one LIGO  instrument and at $1$ Mpc for Advanced LIGO (s15g) and $100$ kpc (s15h), respectively. The abbreviation 
$dist.$ indicates the distance between detector and source, measured in kpc. $f_{c}$ and $h_{c}$ denote the characteristical frequency and amplitude; SNR abbreviates 'signal-to-noise' ratio. We determined the signal-to-noise ratios under the assumption of optimal orientation from detector and source for polar (I) and equatorial (II) emission directions, denoted by \textit{dir}.}
}
\centering
\label{table:2}
\begin{tabular}{crcccccccc}
\hline\hline
\\
Model & dist.[kpc] & dir. & $f_{c}[Hz]$ & $h_{c}$ & SNR \\

\hline
\\
$s15g_{LIGO}$& 10 & I & 524 & $5.1\times10^{-20}$ & 38.4      \\ 
               & & II& 469 & $6.7\times10^{-20}$ & 59.1    \\
\hline
\\
$s15h_{LIGO}$ & 10 & I & 227 & $1.4\times10^{-21}$ & 3.2 \\
               & & II & 165 & $2.1\times10^{-21}$ & 5.6  \\
\hline
\\
$s15g_{AdvLIGO, burst}$& 1000 & I & 812 & $7.6\times10^{-22}$ & 7.3  \\
                     &  & II& 841 & $1.1\times10^{-21}$ & 10.2  \\
\hline
\\
$s15h_{AdvLIGO, nsns}$ & 100 &I & 221 & $ 1.5\times10^{-22}$ & 3.9  \\
                     & & II& 173 & $2.5\times10^{-22}$ & 6.6  \\
\hline
\end{tabular}
\end{table}
The computed SNRs, the characteristical frequencies $f_{c}$ and amplitudes $h_{c}$ are displayed in Table \ref{table:2}. 
The results indicate a fair chance of detecting model s15g within our Galaxy at the present performance of LIGO.
Note that the prospect of observing a narrow band and long-lasting GW signal from the nonaxisymmetric dynamical instability (s15g)
is enhanced, increasing the detection-limit for a facility running at current sensitivity up to about 100kpc. Furthermore, Fig. \ref{fig8.eps} reveals that the signal-to-noise ratio of the nearly non-rotating supernova model s15h, where the bounce signal is only marginally present, is most probably to small for being traced by LIGO, although the peak sensitivity of current detectors lies in the frequency range our results show is dominated by convection. 
The increased detector sensitivities of Advanced LIGO should allow the catching of events such as the rotational core collapse model s15g at distances around $1$ Mpc and the nearly spherically symmetric s15h at about $100$ kpc.

\section{Uncertainties in the underlying models}

Many branches of physics contribute to the models that enable the
prediction of GW signals from core collapse supernovae. Therefore,
it is a difficult task to objectively judge on the relative importance
of uncertainties in GW signal templates. Perhaps we can start by distinguishing
features that appear in generic models from features that appear in
deterministic models. A feature of generic models would be a feature
that appears for all representative models out of a given class, while
a feature of a deterministic model would only appear in a specific
simulation. We believe that most current supernova models should be
seen as generic models that aim to represent a class of stars, rotation
rates, etc. instead of a specific event. Hence, we should concentrate
on the generic features of the predicted GW signals and defer the
discussion of deterministic features to future improved supernova
models. Among the generic features we can further distinguish qualitative
statements and quantitative statements. For example, the finding that
the A\( _{+II} \) quadrupole from a rotating collapse model shows
a pre-bounce rise before a large negative peak as shown in Fig. 3
is a generic qualitative statement \citep{Dimmelmeier07PhysRev},
while the information that the Fourier transform peaks at \( 893 \)
Hz is a quantitative statement.

It is almost impossible to draw the borderline between generic and
deterministic features based on a single simulation. But even if a
full series of models is available, it is difficult to exclude that
some generic features of this series are not influenced by a generic
limitation of the underlying numerical algorithms. Hence, it is very
important that different groups do not just develop the one and best-suited
code for a given problem, but rather a variety of different numerical
approaches so that the results can be compared in order to reveal
the common generic features of the GW signal from core-collapse simulations.

The last stage of the evolution of a massive star proceeds through
very different phases. First, there is the stellar core collapse and
bounce at nuclear density. Then there is a possibly extended accretion
phase that eventually leads to the supernova explosion. These two
phases involve different conditions of matter and pose different challenges
to the numerical modelling. The quality and reliability of the models
depends very strongly on the investigated phase. In the following,
we discuss the uncertainties of the models for each phase separately.

\subsection{Core-collapse and bounce}

The models of stellar core collapse and bounce provide a link between
progenitor star properties and the first strong emergence of a GW
signal. Many previous studies analysed the GW signal based on idealised
input physics in order to study the qualitative dependence of the
core-bounce signal on progenitor star properties (e.g. \citet{1997A&A...320..209Z}).
For example, polytropic equations of state are used and the influence
of neutrino interactions on the collapse dynamics are ignored. This
approach produced a variety of GW bounce signals as a function of
parameters like the angular momentum profile of the progenitor star,
the adiabatic index of the equation of state, or the presence of strong
magnetic fields. As there is no GW emission in spherical symmetry,
these studies have always been conducted in a multi-dimensional setting.

However, the input physics relevant during the collapse of the stellar
core is very rich on the microscopic level \citep{Martinez-Pinedo07}
and this dynamical phase is very sensitive to small perturbations.
The evolution of the collapsing core depends significantly on the
adiabatic index of the equation of state, general relativistic effects,
the electron capture rates on free protons and nuclei, and the coherent
scattering opacities of neutrinos off the different nuclei. Input
physics improvements have been explored in the context of a long-term
modelling effort to clarify the supernova explosion mechanism in spherically
symmetric models. Through the ever-increasing power of computers,
these improvements can now be carried over to the multi-dimensional
predictions of GW signals \citep{M2004}. Few groups
have implemented equations of state that contain nuclear input physics
and added our simple and efficient neutrino physics parametrisation
scheme to get a more accurate mapping of the progenitor properties
onto the expected GW signal from the bounce at nuclear matter densities
\citep{Dimmelmeier07PhysRev,Cerda-Duran2007}.
It was possible to show by two- and three-dimensional general relativistic
simulations that only type I signals are expected to 
occur \citep{Dimmelmeier07PhysRev,OttPhysRev07},
which we independently confirm by the models presented in this paper.

On the other hand, the uncertainties in the progenitor properties
and the equation of state above nuclear density are rather large.
Hence, even with the most accurate numerical scheme it is not possible
to calculate a definitive GW signal from core-bounce. The improved
models of the latest generation help to accurately map progenitor
properties and equation of state properties to a potentially observable
GW signal. This may lead to constraints from future GW observations.
For example, the amplitude and timing of the GW signal with respect
to the neutrino signal strongly depends on the rotation rate of the
inner stellar core. It could also depend on the size of the collapsing
core (weak interactions) and on asymmetric perturbations induced by
convection in the stellar envelope at the time of collapse. The peak
frequency of the Fourier transformed signal contains information on
the compressibility of nuclear matter at bounce.
Hence, it should be possible to compare GW wave signals for different
physical equations of state. One should also investigate a potential
impact of strong magnetic fields on the asymmetries of the collapse
that might become visible in the GW signal \citep{Kotake2004,Obergaulinger2006}.

\subsection{Postbounce phase}

The homologously collapsed stellar core forms a neutron star before
it probably enters an extended accretion phase. During this postbounce
phase layers from the outside of the original iron core fall into
the standing accretion shock that results from core-bounce. They are
shock heated and dissociated, and settle on the protoneutron star.
For several reasons, this phase is much more difficult to accurately
capture in a numerical model than the collapse phase. One problem
is that the neutrinos do not only stream away from the surface of
the neutron star, their fractional absorption behind the standing
accretion shock leads to an essential feedback to the accretion dynamics.
The hot layers around the protoneutron star show several three-dimensional
fluid instabilities that make an accurate treatment of the radiative
coupling between the deleptonisation of the protoneutron star and
the accreting layers very delicate. Additionally, one has to keep
in mind that the density contrast between the center of the neutron
star and these hot layers behind the standing accretion shock may
easily exceed five orders of magnitude, which imposes severe time
step constraints on codes that treat the whole domain consistently
for the extended evolution time until the supernova explosion is thought
to be launched. As the time step of the hydrodynamics part is limited
by the ratio of the zone width to the signal speed, larger time steps
can be taken if the resolution is low in the regions of largest sound
speed. As described above, we can currently perform simulations with
an equidistant resolution of \( 1 \) km. This resolution is conveniently
high in the outer layers behind the standing accretion shock, but
rather low at the surface and in the interior of the protoneutron
star. We avoid numerical artifacts from the steep density gradient
at the surface of the protoneutron star by analytically considering
the hydrostatic density gradient in the TVD advection scheme. However,
the low resolution might still suppress an efficient coupling of low
protoneutron star modes to higher dynamical modes \citep{Weinberg}.
A more flexible grid for our code is under development in order to
adapt the resolution to the local conditions.

For these reasons, we consider the technical uncertainties in the
postbounce phase to be on a similar level than the uncertainties of
the input physics, which of course continue to be present after core-bounce.
Only very few studies have predicted gravitational waves based on
a postbounce model that includes sophisticated neutrino physics in
axisymmetry (e.g. \citet{M2004}). Neutrino physics
approximations have been used in earlier three-dimensional models
e.g. \citet{Fryer2004}. In our current models, we continue
the hydrodynamical simulation to the postbounce phase, but are aware
that the neutrino physics parametrisation cannot handle the neutronisation
burst that sets in shortly after bounce and it does not feature the
deep electron fraction trough that develops behind the shock due to
the copious electron captures. Furthermore, the neutrino parametrisation
scheme does not feature any neutrino heating. Hence, it is important
to distinguish the results from the collapse phase and bounce, where
we believe that the neutrino physics parametrisation is a viable
and reasonably accurate approach, from the dynamics of the postbounce
phase, where the neutrino physics parametrisation is not sufficiently
accurate. The data from the postbounce evolution in our models has
to be understood as an idealised exploration of potential GW features
after the bounce signal. This is currently the state-of-the-art in
three-dimensional simulations and handled the same way in the models
of \citet{OttPhysRev07}.

In order to improve the models also in the postbounce phase, we have
developed the isotropic diffusion source approximation (IDSA) for
the neutrino transport \citep{Liebendoerfer07}
for future models. The scheme has been implemented and tested in spherical
symmetry, but is not yet fully functional in three dimensions. From
basic considerations and comparison to the results in axisymmetry
with accurate neutrino transport \citep{M2004} we
expect that the additional neutrino emission will lead to a more compact
neutron star that is accreting matter from an envelope with stronger
fluid asymmetries. Compared to our current simulations, one could
expect a stronger GW signal in the late postbounce phase with a shorter
periodicity in the signal for comparable initial rotation rates of
the progenitor. However, this will be explored in more detail with
the next generation of postbounce supernova models.


\section{Conclusion}

In this paper we presented the GW analysis of two 3D MHD core collapse
supernova simulations, which differ only in the amount of initial
rotation. Model s15h started with an angular velocity of \( \Omega =0.3 \)
rad/s, model s15g with \( \Omega =2\pi  \) rad/s. We incorporated
progenitor stars from stellar evolution calculations, spherically
symmetric GR effects, the Lattimer-Swesty EoS, magnetic fields and
a neutrino parametrisation scheme that is accurate until the first
few ms after bounce. The particular choice of the initial angular
momentum allowed clear distinctions between gravitational wave features
that stem from rotation or nonaxisymmetric motions.

The amplitudes for the three directions and polarisations, \( A_{+I},A_{\times I},A_{\times II} \),
are not very sensitive to rotation and show a similar size of several
centimetres. Apparently, they initially couple only weakly to rotationally
induced large scale asymmetries in the mass-energy distribution. The
only GW amplitude that is strongly correlated to axisymmetric rotation
at the time of core-bounce turns out to be \( A_{+II} \) in the \( \theta =\pi /2 \),
\( \phi =0 \)-direction. In model s15g, it exceeds all other amplitudes
and in particular the corresponding one from model s15h by more than
one order of magnitude. It shows a clear type I characteristics and
implies that a rotational core collapse stays axisymmetric in the
early postbounce phase, as lately discussed in \citet{2007CQGra..24..139O}.
At the bounce stage of the simulation s15g, the dominant band of emission
is peaking around 893 Hz (s15g), which is in good agreement with the
recent findings from \citet{M2004}, but roughly \( 150 \) Hz higher
than in \citep{Dimmelmeier07PhysRev,2007CQGra..24..139O}. Within the
investigated time window of about \( 70 \) ms duration, the channel
\( A_{+II} \) in model 15g accounts during the bounce and ring-down
phases for \( \approx90 \% \) of the total energy release in GW emission.

The models of the later postbounce phase are yet affected by significant
technical and physical uncertainties. They should be understood as
a qualitative outlook to future models that will have to include better
neutrino transport and a higher resolution of the protoneutron star.
In the postbounce phase, nonaxisymmetric dynamics starts to play an
important role. The wave trains are revived either through convective
or through nonaxisymmetric instabilities in the protoneutron star.
The GW amplitudes from both models show a sustained signal of approximately
constant size, but of different physical origin. In model s15g, the
occurrence of a so-called low \( T/|W| \) instability of dominant
\( m=2 \) character around \( 20 \) ms post-bounce leads to the
prolonged narrow band GW emission at a frequency of \( 905 \) Hz,
which is in good agreement with \citet{2007CQGra..24..139O} in all
points, except their dominant mode is \( m=1 \). On the other hand,
the onset of nonaxisymmetric dynamics in model s15h is triggered by
convective motions due to a negative entropy gradient. In our current
models, the characteristic frequency of emission linked to convective
features takes place below 250 Hz. This number may require adaption
after the inclusion of more accurate neutrino transport. Further,
no sign of a decay, be it due to a dynamical instability or convection,
is present in the long term evolution of the wave trains, as it was
already observed in \citep{M2004,2007CQGra..24..139O}.

Finally we conclude that gravitational waves from the discussed dynamical
features, namely the rotational core bounce and its subsequent ringdown
and low \( T/|W| \) instability related to a core collapse supernova
like model s15g could possibly be detected by a current LIGO instrument
within a distance of \( 10 \) kpc. For model s15h it seems more likely
that LIGO would only catch signals from later stages of the supernova
evolution, meaning frequency distributions caused by convective motion.
The future improvement of detector sensitivities in combination with
several adjustable frequency responses of Advanced LIGO will give
the opportunity to tune the instruments to a particular event/source
and therefore allow to look much deeper into space. This will enhance
the chance of observing core collapse supernovae similar to our models
up to distances of \( 1 \) Mpc.


\begin{acknowledgements}
We would like to thank R. K\"appeli and F.-K. Thielemann for supporting this work in numerous ways. Further acknowledgements go to H. Dimmelmeier and C. D. Ott for enlightening discussions and detailed comments; D. Shoemaker for his very helpful notes on LIGO. The simulations have been carried out on the McKenzie cluster at CITA \citep{Dubinski2003}, the Athena cluster at the University of Basel and the Swiss Supercomputing center CSCS. We acknowledge support by the Swiss National Science Foundation under
grant No. 200020-105328/1 and PP002-106627/1.
\end{acknowledgements}


\end{document}